\documentclass[journal]{vgtc}                     

\onlineid{0}
\usepackage[table,HTML,svgnames]{xcolor}
\definecolor{NavyBlue}{RGB}{0,0,128}
\usepackage{multirow}
\usepackage{booktabs}
\usepackage{siunitx}

\vgtccategory{Research}

\vgtcpapertype{please specify}

\title{Charts-of-Thought: Enhancing LLM Visualization Literacy Through Structured Data Extraction}

\author{
  \authororcid{Amit Kumar Das}{0000-0002-2600-8321},
  \authororcid{Mohammad Tarun}{0009-0001-1708-9924},
  \authororcid{Klaus Mueller}{0000-0002-0996-8590}
}

  \authorfooter{
    \item
    	Amit Kumar Das and Klaus Mueller are with the Computer Science Department, Stony Brook University, USA (E-mails: amitkumar.das@stonybrook.edu, mueller@cs.stonybrook.edu).

        \item
    	Mohammad Tarun is with East West University, Bangladesh (e-mail: md.tarun005@gmail.com).

  }

\abstract{%
  This paper evaluates the visualization literacy of modern Large Language Models (LLMs) and introduces a novel prompting technique called Charts-of-Thought. We tested three state-of-the-art LLMs (Claude-3.7-sonnet, GPT-4.5-preview, and Gemini-2.0-pro) on the Visualization Literacy Assessment Test (VLAT) using standard prompts and our structured approach. The Charts-of-Thought method guides LLMs through a systematic data extraction, verification, and analysis process before answering visualization questions. Our results show Claude-3.7-sonnet achieved a score of 50.17 using this method, far exceeding the human baseline of 28.82. This approach improved performance across all models, with score increases of 21.8\% for GPT-4.5, 9.4\% for Gemini-2.0, and 13.5\% for Claude-3.7 compared to standard prompting. The performance gains were consistent across original and modified VLAT charts, with Claude correctly answering 100\% of questions for several chart types that previously challenged LLMs. Our study reveals that modern multimodal LLMs can surpass human performance on visualization literacy tasks when given the proper analytical framework. These findings establish a new benchmark for LLM visualization literacy and demonstrate the importance of structured prompting strategies for complex visual interpretation tasks. Beyond improving LLM visualization literacy, Charts-of-Thought could also enhance the accessibility of visualizations, potentially benefiting individuals with visual impairments or lower visualization literacy.
}


\graphicspath{{figs/}{figures/}{pictures/}{images/}{./}} 

\usepackage{tabu}                      
\usepackage{booktabs}                  
\usepackage{lipsum}                    
\usepackage{mwe}                       

\usepackage{mathptmx}                  

\begin{document}


\firstsection{Introduction}

\maketitle

As Generative Pre-trained Transformers (GPT) and other Large Language Models (LLMs) continue to advance \cite{IN35}, their impact on the field of visualization is becoming increasingly significant. Previous research has explored implementing LLMs in diverse visualization tasks, including visualization creation \cite{IN16,IN34}, data encoding \cite{IN36}, caption generation \cite{IN26}, design advice \cite{IN23}, and visualization education \cite{IN11}. Researchers have also begun investigating whether LLMs can evaluate visualizations \cite{IN1}. Such evaluations could support visualization sanity checks and provide feedback on visualization readability \cite{IN8}. Traditionally, evaluating visualizations requires significant human effort. If LLMs could support visualization experts in evaluating visual representations at a level comparable to humans, they could offer substantial time and cost savings \cite{IN18}.
\par However, before we can reliably use LLMs to evaluate visualizations, we must ensure these models possess adequate visualization literacy—the ability to read, understand, and interpret visual representations \cite{IN40}. Recent studies have raised important questions about LLMs' visualization literacy capabilities. Hong et al. \cite{IN15} found that both GPT-4 and Gemini failed to achieve the same levels of visualization literacy as the general public when tested on a modified Visualization Literacy Assessment Test (VLAT). Similarly, Bendeck and Stasko \cite{IN2} reported that GPT-4 scored in the 16th percentile of humans on the VLAT, demonstrating particular weakness in value retrieval tasks.
\par These findings contrast the increasing capabilities of multimodal LLMs in other domains \cite{IN6,IN15,IN54,IN55}. We hypothesized that the underperformance in visualization literacy tasks might stem not from inherent limitations in the models but from how they approach the task. When humans interpret visualizations, they often engage in a structured analytical process: first identifying axes and data points, then extracting relevant values, and finally performing calculations or comparisons to answer questions \cite{IN40,IN7}. Standard prompting techniques for LLMs typically do not guide models through such a systematic approach.
\par To address this issue, we introduce a novel prompting strategy called "Charts-of-Thought." This approach is inspired by the Chain-of-Thought prompting technique \cite{IN45}, which has proven effective in improving LLMs' reasoning capabilities for complex mathematical and logical tasks. Our Charts-of-Thought method guides LLMs through a structured data extraction, verification, and analysis process before answering visualization-related questions.
\par Our Charts-of-Thought approach builds on established theories of how humans process visualizations. Bertin's work on "reading levels" \cite{X1} describes how viewers progress from identifying elements to extracting meaning—a process our structured prompts mirror. Cleveland and McGill \cite{Y1} showed that humans follow a sequence when decoding visualizations: they extract values before making comparisons. Carpenter and Shah \cite{Z1} proposed that graph comprehension involves cycles of pattern recognition, interpretation, and integration with prior knowledge. The four tasks in our method (data extraction, sorting, verification, and analysis) align with how people process visualizations step by step rather than all at once, as documented by Hegarty \cite{W1}. By structuring prompts to follow this natural cognitive progression, we help LLMs overcome the visualization literacy challenges identified in previous research.

\par Using this approach, we evaluate three state-of-the-art multimodal LLMs—Claude-3.7-sonnet, GPT-4.5-preview, and Gemini-2.0-pro—on the Modified VLAT and VLAT \cite{IN40}, a standardized test for assessing visualization literacy. We compare their performance using both standard prompting techniques and our Charts-of-Thought method and examine how well these models perform across different visualization types, task categories, and question difficulty levels.
\par Through this comprehensive evaluation framework, we aim to address several key questions about the capabilities and limitations of modern LLMs in visualization interpretation tasks:
\begin{itemize} 
    \item \textbf{\textit{RQ1:}} 
    Can modern multimodal LLMs achieve human-level visualization literacy when guided through a structured analytical process?
    \item \textbf{\textit{RQ2:}} 
    How does the Charts-of-Thought prompting strategy affect LLM performance on visualization literacy tasks compared to standard prompting?
    \item \textbf{\textit{RQ3:}} 
    How do different LLMs perform across various visualization types and analytical tasks?
    \item \textbf{\textit{RQ4:}} 
    To what extent do LLMs rely on their prior knowledge versus information in the visualization when answering questions?    

\end{itemize}

\par Our findings reveal that when properly guided, LLMs can not only match but exceed human performance on standardized visualization literacy assessments. These results challenge current assumptions about LLMs' capabilities and suggest new possibilities for their application in visualization contexts. Beyond raw performance metrics, our analysis uncovers patterns in how different models handle various visualization types and analytical tasks, providing practical insights for researchers and practitioners looking to leverage these technologies.
\par In summary, the main contributions of this work are as follows:

\begin{itemize} 
    \item
    We introduce Charts-of-Thought, a structured prompting technique that guides LLMs through systematic data extraction, verification, and analysis processes for visualization interpretation.

    \item
   We demonstrate that Claude-3.7-sonnet, when using our approach, achieves a VLAT score of 50.17, substantially exceeding the human mean of 28.82 and challenging the view that current LLMs lack sufficient visualization literacy.

    \item
   We show that our prompting strategy consistently improves performance across all tested models, with score increases of 21.8\% for GPT-4.5-preview, 9.4\% for Gemini-2.0-pro, and 13.5\% for Claude-3.7-sonnet compared to standard prompting.

    \item
   We provide a detailed analysis of LLM performance across different visualization types and tasks, revealing patterns of strength and weakness to inform future applications.
   \item
   We propose a framework to test and enhance LLM visualization literacy for future research and applications.

\end{itemize}


In the following, Section 2 presents related work, Section 3 presents our Charts-of-Thought structured prompting approach and contrasts it with conventional prompting, Sections 4-6 present evaluations with a modified VLAT, the original VLAT, and a chart question answering system, and Section 7 and 8 present a discussion and conclusions.

\section{Related Work}
This section presents relevant prior work in visualization literacy, chart question-answering systems, and the application of Large Language Models (LLMs) in visualization research. We also discuss existing approaches to evaluating LLMs on domain-specific tasks.
\subsection{Visualization Literacy}
As data and data visualizations become increasingly prevalent across domains \cite{IN2}, the ability to read and interpret visualizations—visualization literacy—has emerged as a crucial skill \cite{IN40, IN7}. Börner et al. \cite{IN7} define visualization literacy as "the ability and skill to read and interpret visually represented data in and to extract information from data visualizations." Over the past decade, researchers have developed various approaches to both assess \cite{IN7,IN8,IN40} and improve \cite{IN36,IN61} human visualization literacy capabilities.
\par The Visualization Literacy Assessment Test (VLAT) \cite{IN40} stands as one of the most comprehensive and widely used tools for measuring visualization literacy. Developed by Lee et al., the VLAT consists of 53 multiple-choice questions spanning 12 common visualization types. The test covers a range of analytical tasks previously identified as fundamental in data visualization work \cite{IN3,IN9}, including retrieving values, finding extrema, determining ranges, identifying trends, making comparisons, finding anomalies, identifying clusters, and recognizing hierarchical structures. Lee et al. found that the VLAT effectively distinguishes between individuals with different levels of visualization experience and education.
\par Building on this work, Pandey and Ottley \cite{IN31} developed a condensed version called the Mini-VLAT, which contains 12 items for a more efficient assessment of visualization literacy. Other researchers have investigated visualization literacy from different perspectives, including the ability to interpret unfamiliar visualization techniques \cite{IN7} and the impact of visual embellishments on comprehension \cite{IN4,IN42}.
\par While these tools have been extensively used to assess human visualization literacy, until recently, they have not been applied to evaluate the visualization literacy of LLMs. Recent work by Hong et al. \cite{IN15} and Bendeck and Stasko \cite{IN2} has begun to address this gap by evaluating the performance of multimodal LLMs on visualization literacy tasks. Both studies found that current LLMs generally underperform compared to humans on standard visualization literacy assessments, with particular difficulties in tasks requiring precise value retrieval and color interpretation. Our work builds upon these findings while exploring whether structured prompting strategies can overcome these limitations.
\subsection{Chart Question Answering}
Chart Question Answering (CQA) represents another approach to evaluating the ability of systems to interpret visualizations. CQA research focuses on developing systems that can automatically answer natural language questions about charts \cite{IN28}. This field has seen significant advances in recent years, with researchers developing various approaches to tackle this challenge.
\par Early CQA datasets such as DVQA \cite{IN20} and FigureQA \cite{IN20} introduced synthetic charts paired with questions to test different aspects of chart understanding. Kafle et al. \cite{IN21} proposed the PReFIL algorithm for answering chart-based questions, which integrates parallel processing of questions and images. Later work by Masry et al. \cite{IN29} introduced ChartQA, a comprehensive dataset featuring both human- and machine-generated question-answer pairs, providing a more realistic evaluation benchmark. Recent work models human behavior during the interpretation of charts. Chartists \cite{Chartist} demonstrate that humans employ systematic eye movements for various tasks, including value retrieval and identifying extremes. SalChartQA \cite{SalChartQA} reveals that question types drive where users focus their attention on charts. Both studies confirm that humans follow structured processes: they identify relevant elements, extract values, and then perform analysis. Our Charts-of-Thought approach mirrors these human behavioral patterns in LLM prompts.
\par Among these systems, Kim et al.'s approach \cite{IN32} is particularly relevant to our work. Their system consists of three stages: (1) data extraction, (2) question processing and answering, and (3) explanation generation. Bendeck and Stasko \cite{IN2} found that GPT-4, when provided with the underlying data, achieved 87\% accuracy on the CQA dataset from Kim et al., outperforming their system significantly. However, without access to the data, GPT-4's performance dropped to 31\%, highlighting the importance of effective data extraction for visualization interpretation.
\par Our Charts-of-Thought approach draws inspiration from Kim et al.'s pipeline, but instead of relying on separate components for data extraction, question answering, and explanation generation, we guide LLMs to perform these steps sequentially within a single prompt. This approach leverages the end-to-end capabilities of modern multimodal LLMs while ensuring they follow a structured analytical process.

\begin{figure*}[tbp]
  \centering
  \begin{subfigure}[b]{0.22\textwidth}
    \centering
    \includegraphics[width=\textwidth]{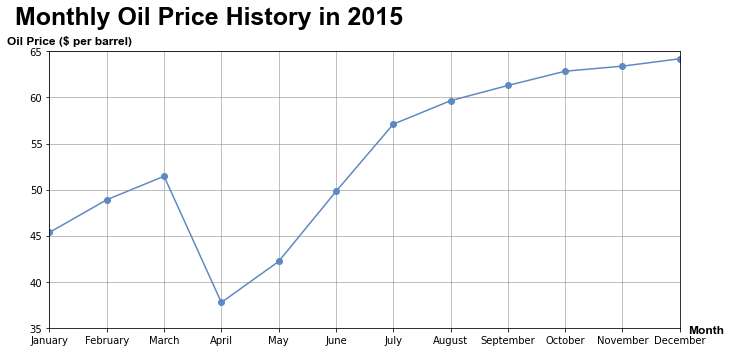}
    \caption{Line Chart}
    \label{fig:1a}
  \end{subfigure}%
  \hfill%
  \begin{subfigure}[b]{0.22\textwidth}
    \centering
    \includegraphics[width=\textwidth]{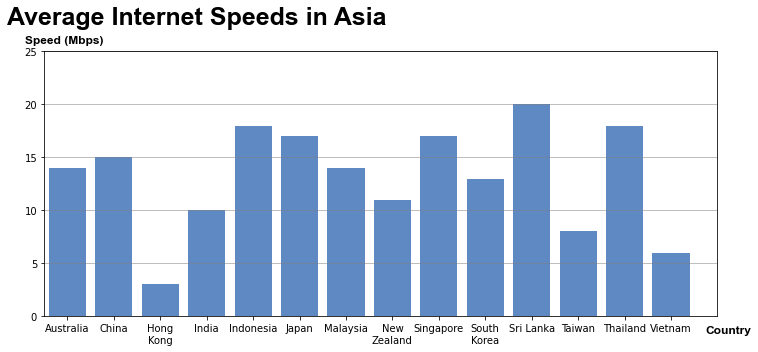}
    \caption{Bar Chart}
    \label{fig:1b}
  \end{subfigure}%
  \hfill%
  \begin{subfigure}[b]{0.22\textwidth}
    \centering
    \includegraphics[width=\textwidth]{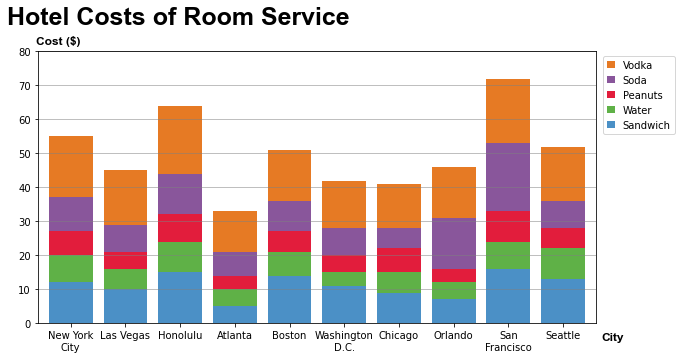}
    \caption{Stacked Bar Chart}
    \label{fig:1c}
  \end{subfigure}%
  \hfill%
  \begin{subfigure}[b]{0.22\textwidth}
    \centering
    \includegraphics[width=\textwidth]{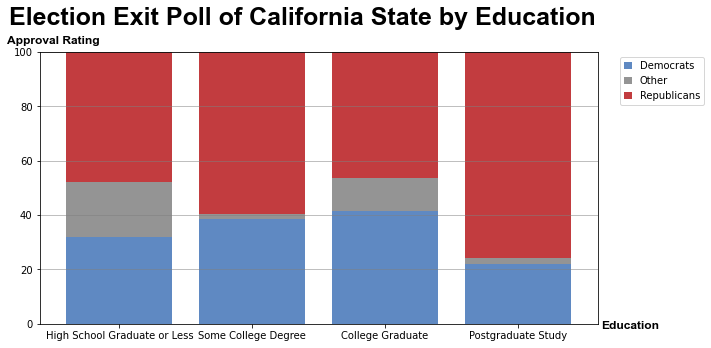}
    \caption{100\% Stacked Bar Chart}
    \label{fig:1d}
  \end{subfigure}

  \vspace{0.3em}
  \begin{subfigure}[b]{0.14\textwidth}
    \centering
    \includegraphics[width=\textwidth]{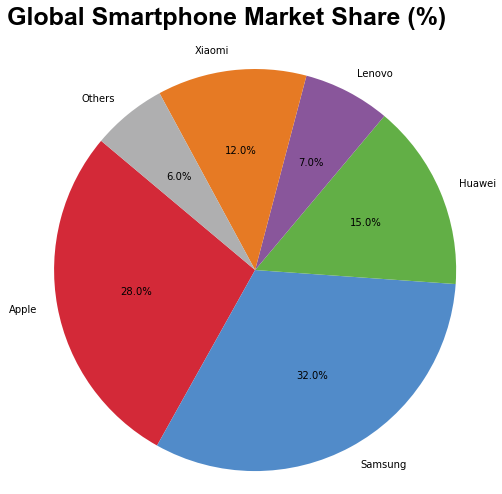}
    \caption{Pie Chart}
    \label{fig:2a}
  \end{subfigure}%
  \hfill%
  \begin{subfigure}[b]{0.22\textwidth}
    \centering
    \includegraphics[width=\textwidth]{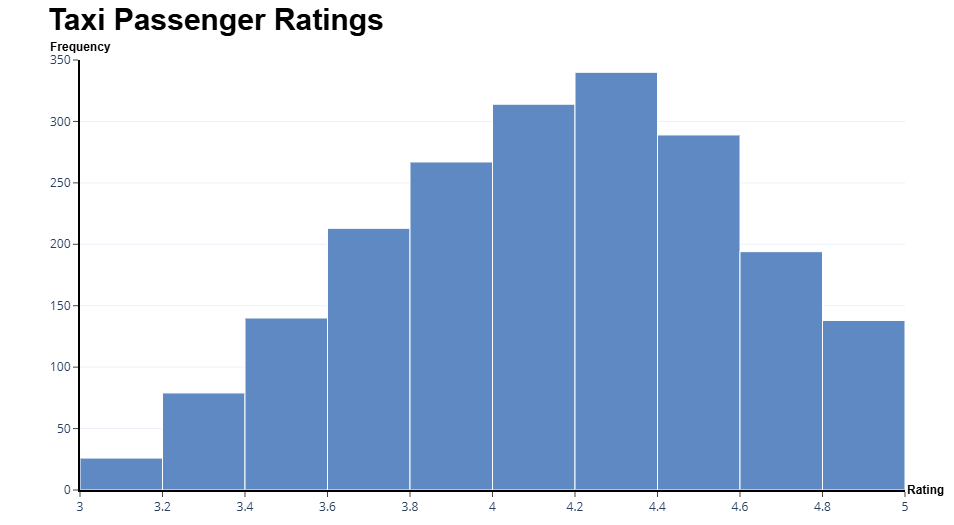}
    \caption{Histogram}
    \label{fig:2b}
  \end{subfigure}%
  \hfill%
  \begin{subfigure}[b]{0.22\textwidth}
    \centering
    \includegraphics[width=\textwidth]{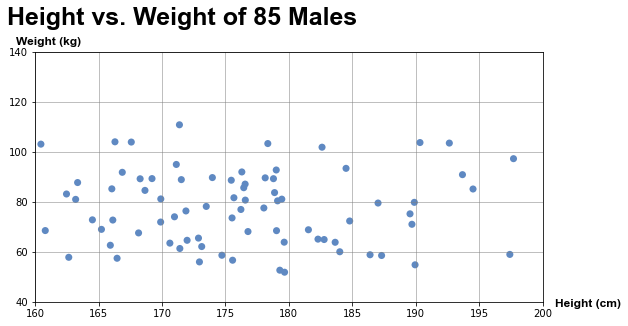}
    \caption{Scatterplot}
    \label{fig:2c}
  \end{subfigure}%
  \hfill%
  \begin{subfigure}[b]{0.22\textwidth}
    \centering
    \includegraphics[width=\textwidth]{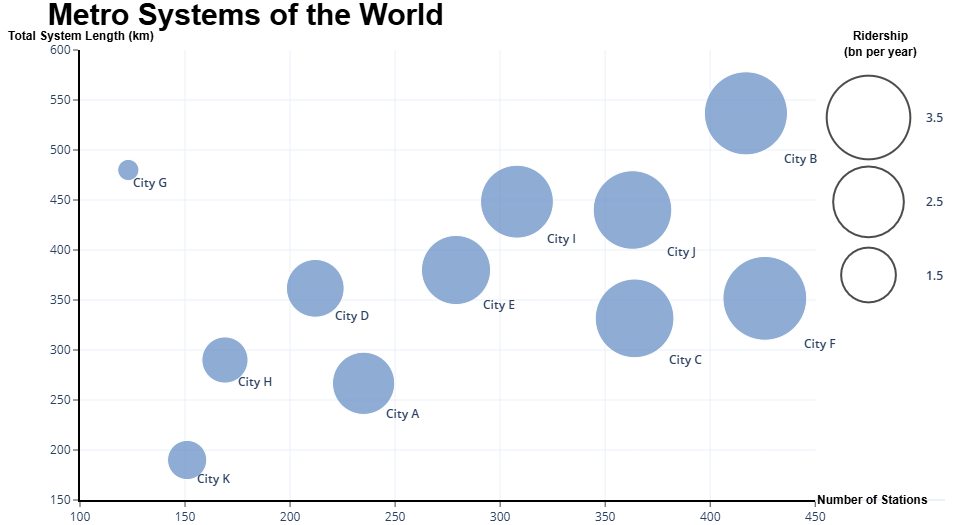}
    \caption{Bubble Chart}
    \label{fig:2d}
  \end{subfigure}

  \vspace{0.3em}
  \begin{subfigure}[b]{0.22\textwidth}
    \centering
    \includegraphics[width=\textwidth]{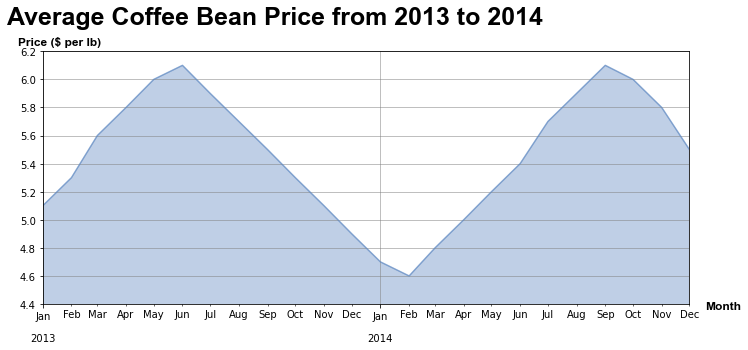}
    \caption{Area Chart}
    \label{fig:3a}
  \end{subfigure}%
  \hfill%
  \begin{subfigure}[b]{0.22\textwidth}
    \centering
    \includegraphics[width=\textwidth]{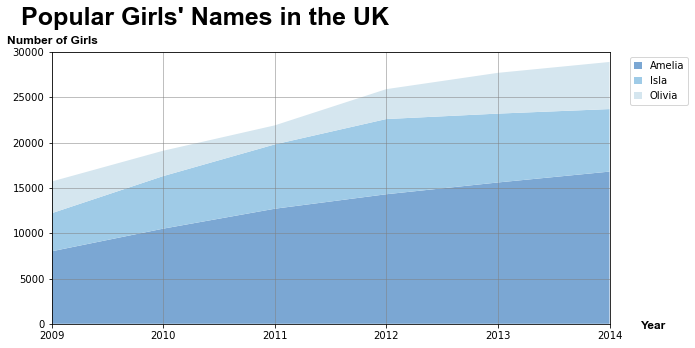}
    \caption{Stacked Area Chart}
    \label{fig:3b}
  \end{subfigure}%
  \hfill%
  \begin{subfigure}[b]{0.22\textwidth}
    \centering
    \includegraphics[width=\textwidth]{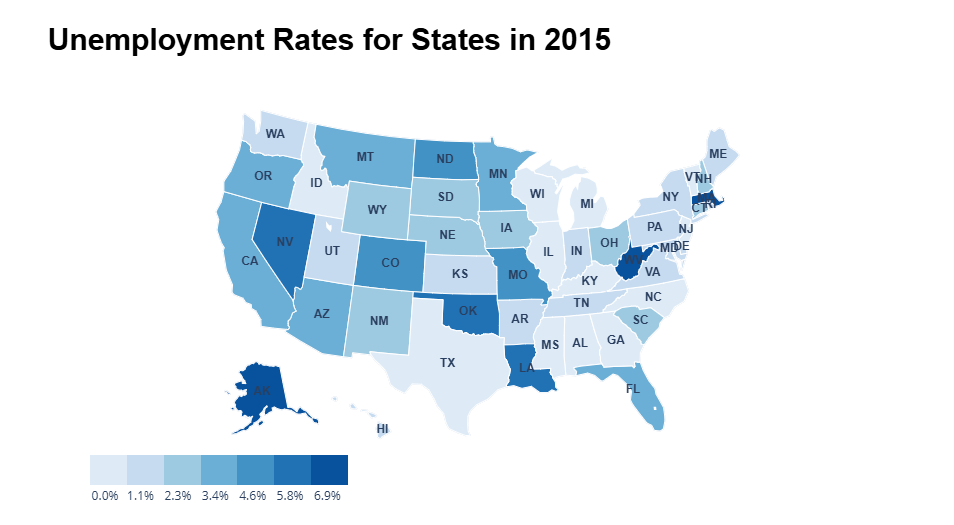}
    \caption{Choropleth Map}
    \label{fig:3c}
  \end{subfigure}%
  \hfill%
  \begin{subfigure}[b]{0.22\textwidth}
    \centering
    \includegraphics[width=\textwidth]{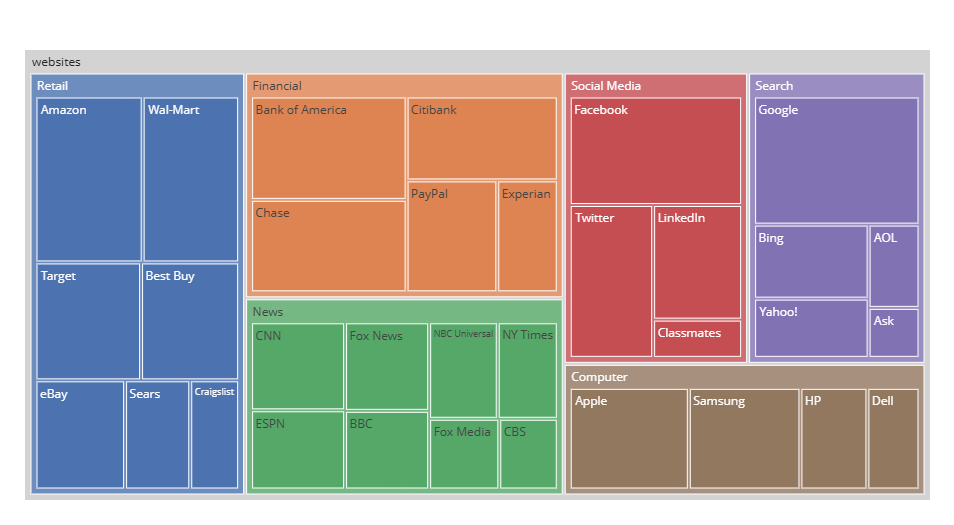}
    \caption{Treemap}
    \label{fig:3d}
  \end{subfigure}
  
  \caption{Complete set of 12 visualization types, recreated from VLAT examples with modified data. These charts represent the full scope of visualization literacy tasks tested, spanning fundamental chart types from simple bar charts to complex treemaps and choropleth maps. Each visualization type was evaluated with 3-8 associated questions to assess different analytical tasks.}
  \label{fig:fig_1}
\end{figure*}

\subsection{LLMs in Visualization Research}
Research at the intersection of LLMs and visualization can be broadly categorized into two areas: visualization for LLMs and LLMs for visualization \cite{IN75}. Visualization for LLMs involves using visualization techniques to help users understand and interact with LLMs, including visualizing model internals \cite{IN20,IN43,IN44}, aiding prompt engineering \cite{IN23,IN67,IN73}, and evaluating model performance \cite{IN16, IN17, IN45}.
\par The second category—LLMs for visualization—focuses on harnessing LLMs to advance visualization research and applications. Researchers have employed LLMs to generate visualization code \cite{IN21,IN50}, create charts directly \cite{IN41,IN77}, generate titles and captions \cite{IN26,IN46}, enhance visualization systems with natural language interfaces \cite{IN53,IN60,IN62,IN64}, and support data-driven storytelling \cite{IN13, IN13, IN69}. Chen et al. \cite{IN11} explored using GPT for a data visualization course, finding that the model could successfully complete diverse visualization tasks but struggled with certain aspects of visualization design and interpretation.
\par Our work contributes to this growing body of research by investigating how LLMs can be effectively prompted to interpret and analyze visualizations. While prior work has primarily focused on using LLMs to create visualizations, our research examines their potential as tools for visualization interpretation and evaluation.
\subsection{Prompting Techniques for LLMs}
The performance of LLMs depends significantly on how they are prompted \cite{IN74}. Researchers have developed various prompting techniques to improve LLM performance across different tasks, with Chain-of-Thought (CoT) prompting \cite{IN45} emerging as one of the most effective approaches for complex reasoning tasks.
\par CoT prompting guides LLMs to break down complex problems into intermediate steps before arriving at a final answer. Wei et al. \cite{IN45} demonstrated that this approach significantly improves LLM performance on arithmetic, commonsense, and symbolic reasoning tasks. Building on this concept, researchers have developed variations such as Tree-of-Thoughts \cite{IN79}, which explores multiple reasoning paths, and Self-Consistency \cite{IN80}, which generates multiple reasoning chains and selects the most consistent answer.
\par In the visualization domain, Feng et al. \cite{INN23} introduced PromptMagician, an interactive prompt engineering tool for text-to-image creation that helps users refine prompts for generating visualizations. However, little work has explored specialized prompting techniques for improving LLMs' ability to interpret visualizations.
\par Our Charts-of-Thought approach adapts the principles of CoT prompting specifically for visualization interpretation tasks. By guiding LLMs through a structured process of data extraction, verification, and analysis, we aim to improve their visualization literacy capabilities in a way that aligns with how humans approach these tasks.
\subsection{Evaluating LLMs on Domain-Specific Tasks}
As LLMs continue to advance, researchers have developed benchmarks to assess their capabilities across various domains. Studies have evaluated LLMs on standardized tests in higher education \cite{IN54}, including the Law School Admission Test (LSAT) \cite{IN15}, the Medical College Admission Test (MCAT) \cite{IN6}, and the U.S. Medical Licensing Examination \cite{IN55}. In the visualization domain, Chen et al. \cite{IN11} found that GPT-4 could score 80\% on quizzes and homework from a data visualization course. More directly relevant to our work, Bendeck and Stasko \cite{IN2} evaluated GPT-4's visualization literacy using the VLAT and several other task sets, finding that the model scored in the 16th percentile of humans. Similarly, Hong et al. \cite{IN15} evaluated GPT-4 and Gemini on a modified VLAT, concluding that both models failed to achieve human-level visualization literacy.
\par For multimodal LLMs specifically, HallusionBench \cite{IN27} introduced a diagnostic suite to evaluate vision-language models on various tasks, including some chart interpretation questions. However, this benchmark focuses primarily on identifying hallucinations rather than comprehensively assessing visualization literacy.
\par Our work builds on these evaluation approaches while introducing a novel prompting strategy designed to enhance LLMs' performance on visualization literacy tasks. By systematically comparing the performance of multiple state-of-the-art LLMs across different prompting conditions, we provide insights into these models' current capabilities and limitations for visualization interpretation tasks.
\section{Structured Prompting with Charts-of-Thought}
To evaluate the visualization literacy of modern LLMs and assess the effectiveness of our Charts-of-Thought prompting approach, we conducted a series of experiments using standardized visualization literacy assessment tasks. This section details our overall methodology, including the LLMs evaluated, our experimental design, and the prompting strategies we compared.
\subsection{Large Language Models Evaluated}
We evaluated three state-of-the-art multimodal LLMs that support image analysis capabilities:
\begin{itemize} 
    \item \textbf{\textit{Claude-3.7-sonnet:}} 
    Anthropic's latest multimodal model (released on February 24, 2025) as of our experiments (March 2025), with a knowledge cutoff date of October 2024.
    \item \textbf{\textit{GPT-4.5-preview:}} 
    OpenAI's most advanced multimodal model (released on February 27, 2025) at the time of our experiments, with a knowledge cutoff date of October 2023.
    \item \textbf{\textit{Gemini-2.0-pro:}} 
    Google's latest multimodal model (released on February 05, 2025) with a knowledge cutoff date of June 2024.  
\end{itemize}
\par We selected these models based on their widespread use and accessibility through their official APIs, which enabled us to conduct controlled, large-scale experiments. All three models support multimodal input, allowing us to provide visualization images directly without relying on textual descriptions \cite{IN23} or SVG code \cite{IN11}. We considered using other multimodal models such as DeepSeek-V3, but at the time of our study, some limitations in their API capabilities (such as lack of support for image uploads through the API) prevented their inclusion.
\subsection{Experimental Design}
To thoroughly assess the visualization literacy of the selected LLMs, we used the Visualization Literacy Assessment Test (VLAT) \cite{IN40} as our primary evaluation framework. The VLAT consists of 53 multiple-choice questions covering 12 
visualization types and 8 analytical tasks.
\par Given that the VLAT was published in 2017, we were concerned that current LLMs might have encountered these materials during their training. We conducted a pilot test using GPT-4.5 on the original VLAT questions to investigate this possibility. Upon analyzing the results, we observed that for certain questions, GPT-4.5 consistently provided the same responses as those in the original VLAT, suggesting potential prior exposure to these materials.
To address this concern, we used two versions of the VLAT for our experiments: \\
\begin{itemize} 
    \item \textbf{\textit{Modified VLAT:}} 
    We recreated all charts from the original VLAT with randomized data values while maintaining the same design style. We manually adjusted the randomized data to ensure that certain task types, such as finding correlations or trends, remained feasible. We also updated the questions and answer choices to reflect the new data, ensuring that our answers differed from those in the original VLAT.
    \item \textbf{\textit{Original VLAT:}} 
    We used the original VLAT charts and questions as published by Lee et al. \cite{IN40} to compare with our modified version. 
\end{itemize}
\par To further address concerns about data contamination, we conducted a detailed analysis of potential prior exposure. We identified specific questions where models showed suspicious performance patterns. For example, GPT-4.5 answered nine questions from the original VLAT with identical wording to the published answers, suggesting memorization rather than reasoning. These questions primarily involved pie charts and simple bar charts with distinctive data patterns. To verify that our modifications were effective, we implemented a three-step verification process. First, we flagged questions where models provided identical responses matching published answers without showing reasoning steps. Second, we analyzed these flagged questions, confirming that memorization occurred most often in charts with distinctive visual signatures. Third, we validated our approach by testing both versions on the same model—when a model performed well on original charts but struggled with our modified versions of similar complexity, we confirmed our modifications prevented data contamination. This process ensured our test measured true visualization literacy rather than memorization and guided our efforts to thoroughly change both the data values and visual patterns in our modified versions.

\par 

\par We verified that the LLMs' knowledge cutoff dates (October 2024 for Claude-3.7-sonnet, June 2024 for Gemini-2.0-pro, and October 2023 for GPT-4.5-preview) were all after the VLAT's publication (2017), but we took precautions to ensure our modified versions were not publicly available online before conducting our experiments in March 2025.
\par In total, our evaluation utilizes two complete datasets: the original VLAT, comprising 53 questions across 12 visualization types, and our modified VLAT, which includes 53 questions with altered data values, questions, and answer choices across 12 recreated visualizations. Each chart has 3-8 associated questions, totaling 53 questions per dataset. Each of the 53 questions was tested three times per LLM model per prompting strategy, resulting in 318 trials per LLM model for each dataset (53 questions $\times$ 3 trials $\times$ 2 prompting strategies), totaling 636 trials per LLM model across both datasets (318 $\times$ 2 datasets). This approach creates functionally independent datasets while maintaining identical analytical complexity. Fig. \ref{fig:fig_1} shows examples of the modified visualizations created for our experiments, which include line charts, bar charts, stacked bar charts, 100\% stacked bar charts, pie charts, histograms, scatterplots, bubble charts, area charts, stacked area charts, choropleth maps, and treemaps.
\subsubsection{Prompting Strategies}
Through our pilot studies, we discovered that different prompting strategies can significantly affect LLMs' performance on visualization literacy tasks. We developed and compared two distinct prompting approaches:
\begin{itemize} 
    \item \textbf{\textit{Generic Prompt (Prompt \#1):}} 
    A straightforward instruction asking the LLM to answer a multiple-choice question based on the provided visualization, following the approach used by Bendeck and Stasko \cite{IN2}. The prompt format was:\\
    \textit{I am about to show you an image and ask you a multiple-choice question about that image. Select the BEST answer based only on the chart and not external knowledge.}
    \item \textbf{\textit{Charts-of-Thought Prompt (Prompt \#2)}} 
    A structured, multi-step prompt designed to guide the LLM through a systematic approach to extracting and analyzing data from the visualization before answering the question. The prompt format was:\\
    \textit{I am about to show you a graph and ask you a multiple-choice question about that graph.\\
    Task 1: Data Extraction and Table Creation: First, explicitly list ALL numerical values you can identify on both axes, then create a structured table using markdown syntax that includes ALL data points you identified above with appropriate column headers with units.\\
    Task 2: Sort the data: Sort the data in descending order by the numerical values.\\
    Task 3: Data Verification and Error Handling: Double-check if your table matches ALL elements in the graph by comparing each value in your table with the graph and updating your table with correct values; verify the sorting is correct, and before proceeding, confirm all corrections have been made and use ONLY the corrected data for analysis.\\
    Task 4: Question Analysis: Using ONLY the verified data in your table, compare EACH value individually with the reference value; for "less than" comparisons, mark ALL values that are even slightly below the reference; for "greater than" comparisons mark ALL values that are even slightly above the reference, and show each comparison on a new line.\\
    Let's solve this step by step.}
\end{itemize}
\par The Charts-of-Thought approach was inspired by the Chain-of-Thought prompting technique \cite{IN45,IN455}, which has been shown to improve LLMs' performance on complex reasoning tasks. Our approach specifically adapts this concept to visualization analysis by breaking down the task into explicit steps of data extraction, verification, and structured analysis. The instruction "list all numerical values on both axes" guides a thorough data extraction process across all chart types. For non-axial visualizations like pie charts, LLMs appropriately extracted relevant values. This prompt structure proved effective across all 12 visualization types tested.

\subsubsection{Example Responses to Different Prompts}
\par Our testing process followed a consistent sequence. First, we provided the LLM with one visualization, one associated question, and the multiple-choice options. Then, we added either the Generic prompt or the Charts-of-Thought prompt to instruct the LLM on how to approach the task. The prompts served as process guidelines for the models, with the Generic prompt offering minimal direction, while the Charts-of-Thought prompt structured the analysis into specific steps. This one-by-one testing method was applied uniformly across all visualization–question pairs in our study.

\par During our prompt development, we tested several variations to identify the most critical components of Charts-of-Thought. The data extraction and verification steps showed the greatest impact on performance. Removing the instruction to "create a structured table" decreased accuracy by 18\%, while the verification step requiring models to "double-check if your table matches ALL elements" improved accuracy by 14\%. The sorting step enhanced performance by 11\% for comparison and trend tasks. Both the specific phrasing and ordering of tasks proved important—rearranging steps (e.g., sorting before extraction) decreased performance by 7\%, suggesting our sequence aligns with effective cognitive processing of visualizations. Among individual steps, data extraction proved most error-prone when visualization complexity increased while sorting showed the highest robustness across different chart types.
\par
\begin{figure}[!thb]
    \centering
    \includegraphics[width=0.49\textwidth, alt={Comparison}]{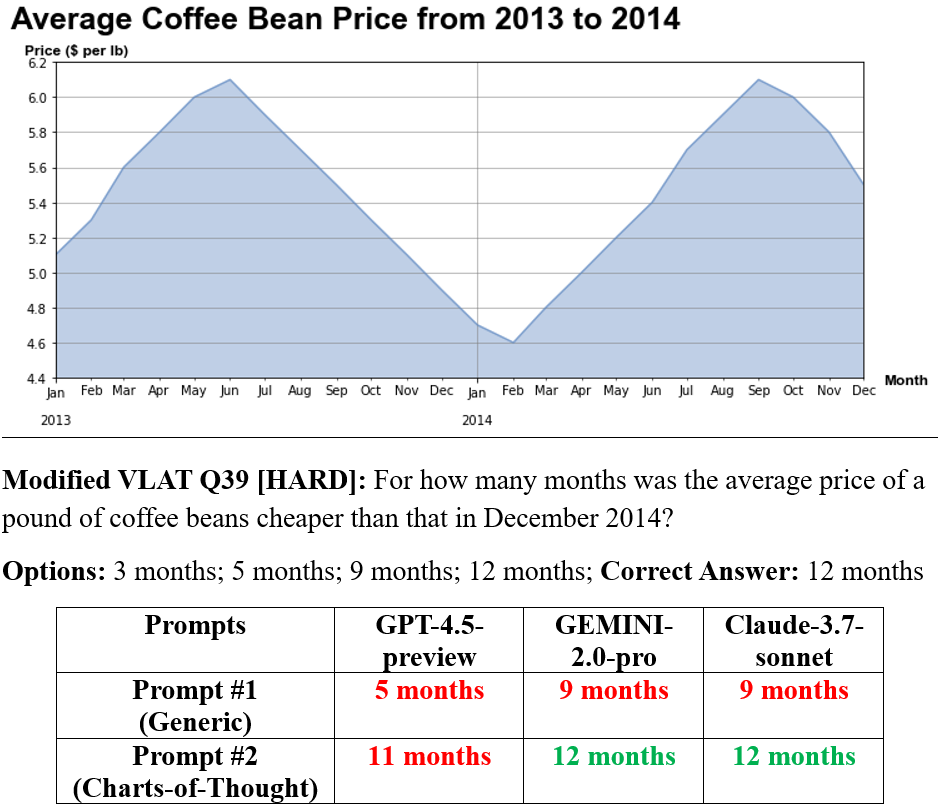}
    \caption{The responses of the three tested LLMs to the modified VLAT Q39 (a hard question) for both the Generic prompt and our Charts-of-Thought prompt. With the Generic prompt, none of the models came even close to the correct answer, while with our Charts-of-Thought prompt, two models returned the correct answer, with the third being close.}
    \label{fig:fig_2}
\end{figure}

 To illustrate how these prompting strategies affect LLM responses, Fig. \ref{fig:fig_2} shows an example visualization used in our study, an example VLAT question, and the responses of the three LLMs we tested to both of these prompts.  

\section{Experiments with the Modified VLAT}
We ran each experimental condition (Generic Prompt and Charts-of-Thought Prompt) on 53 VLAT questions three times per model to account for variance in LLM responses. We recorded the raw scores (number of correct answers) and calculated the VLAT score using the scoring scheme from the paper \cite{IN40}, which includes a guessing penalty.
\par For scoring, we set strict criteria: a model received credit for a correct answer only if it provided the correct response in all three trials. This approach penalized inconsistency and provided a more conservative measure of LLM capabilities.
\subsection{Overall Performance}
Table \ref{tab:1} shows the mean scores for each LLM on the modified VLAT using both prompting strategies. We present both raw scores (number of correctly answered questions out of 53) and scores calculated using the VLAT scoring scheme from the original paper \cite{IN40}, which includes penalties for guessing.

%


\begin{table}[htbp]
\centering
\setlength{\tabcolsep}{4pt}
\caption{VLAT results for modified visualizations with different prompting}
\begin{tabular}{l c c c c}
\toprule
& \multicolumn{2}{c}{\textbf{Generic}} & \multicolumn{2}{c}{\textbf{Charts-of-Thought}} \\
\cmidrule(lr){2-3} \cmidrule(lr){4-5}
\textbf{Model} & \textbf{Mean} & \textbf{VLAT} & \textbf{Mean} & \textbf{VLAT} \\
& \textbf{Raw} & \textbf{Score} & \textbf{Raw} & \textbf{Score} \\
& \textbf{Score} & \textbf{} & \textbf{Score} & \textbf{} \\
\midrule
GPT-4.5 & 34.67 & 23.50 & 42.00 & 35.67 \\
GEMINI-2.0 & 42.67 & 37.72 & 46.67 & 42.78 \\
Claude-3.7 & 44.33 & 39.89 & \textbf{50.33} & \textbf{49.44} \\
\bottomrule
\end{tabular}

\label{tab:1}
\end{table}

\par Using the Charts-of-Thought prompting strategy, Claude-3.7-sonnet achieved the highest score with a mean of 50.33 correct answers and a VLAT score of 49.44. This substantially exceeds the human baseline of 28.82 reported in the original VLAT study. Both GPT-4.5-preview and Gemini-2.0-pro also performed well with Charts-of-Thought prompting, achieving VLAT scores of 35.67 and 42.78, respectively.

\par The results in Table \ref{tab:2} show that the Charts-of-Thought prompting strategy improved performance across all models compared to standard prompting. The most significant improvement occurred with GPT-4.5-preview, which saw a 21.8\% increase in raw score when switching from standard prompting to Charts-of-Thought prompting. Gemini-2.0-pro and Claude-3.7-sonnet showed improvements of 9.4\% and 13.5\%, respectively.

\begin{table}[htbp]
\centering
\setlength{\tabcolsep}{4pt}
\caption{Score Improvement of Generic to Charts-of-Thought prompting}
\renewcommand{\arraystretch}{1.3}
\begin{tabular}{|l|c|p{3.0cm}|}  
\hline
\rowcolor[HTML]{EFEFEF} 
\textbf{LLM Model} & \textbf{Improvement} & \textbf{Visual Comparison} \\
\hline
GPT-4.5-preview & 21.80\% & 
\begin{tikzpicture}[baseline]
\fill[color={rgb:red,16;green,163;blue,127}] (0,0) rectangle (2.18,0.3);  
\node[text=white, right] at (0.3,0.15) {\scriptsize{21.80\%}};
\end{tikzpicture} \\
\hline
GEMINI-2.0-pro & 9.38\% & 
\begin{tikzpicture}[baseline]
\fill[color={rgb:red,255;green,92;blue,138}] (0,0) rectangle (0.94,0.3);  
\node[text=white, right] at (0.3,0.15) {\scriptsize{9.38\%}};
\end{tikzpicture} \\
\hline
Claude-3.7-sonnet & 13.53\% & 
\begin{tikzpicture}[baseline]
\fill[color={rgb:red,233;green,150;blue,122}] (0,0) rectangle (1.35,0.3);  
\node[text=white, right] at (0.3,0.15) {\scriptsize{13.53\%}};
\end{tikzpicture} \\
\hline
\end{tabular}

\label{tab:2}
\end{table}

\subsection{Performance by Question Difficulty}
\begin{figure}[!thb]
    \centering
    \includegraphics[width=0.49\textwidth, alt={Comparison}]{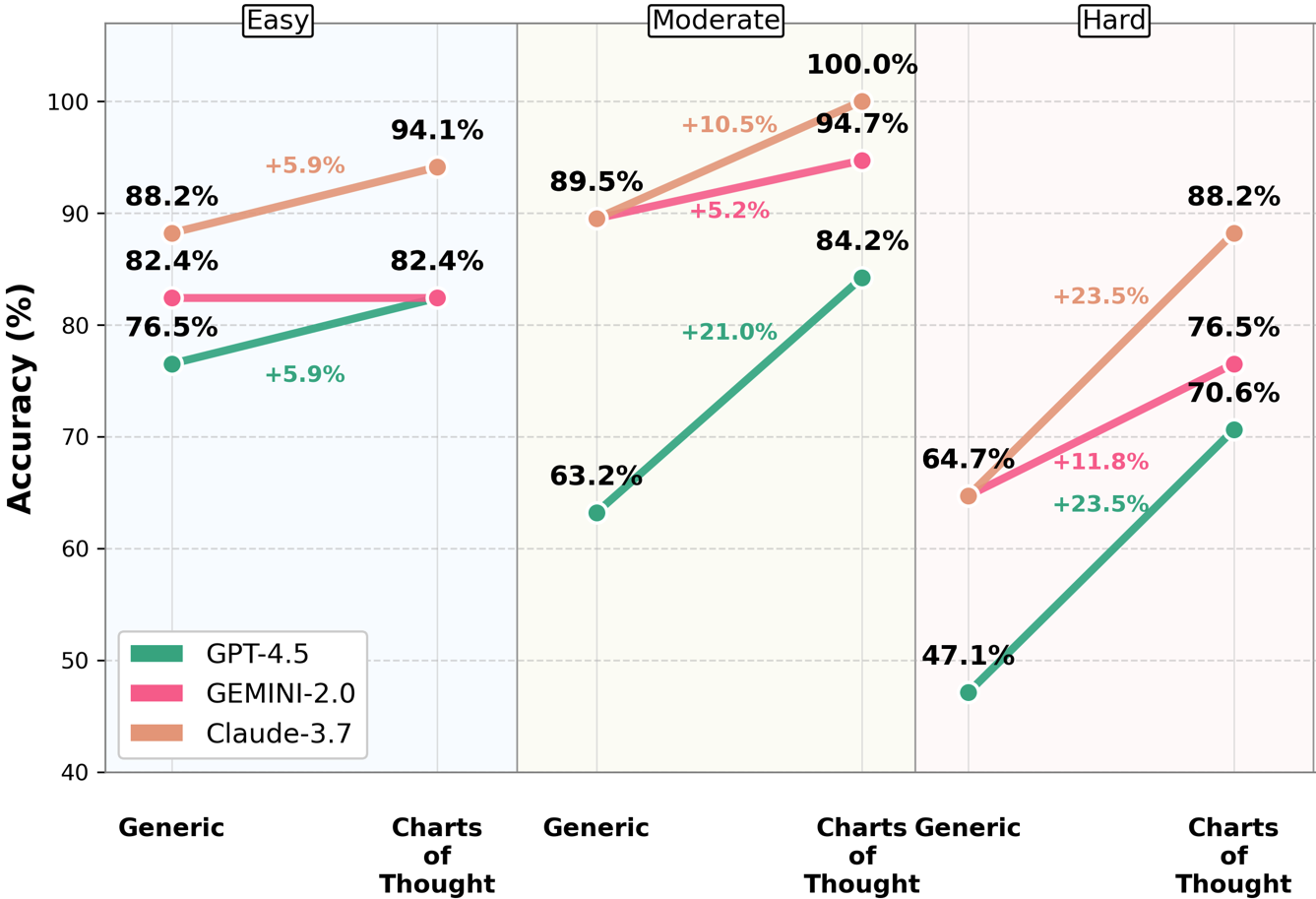}
    \caption{Modified VLAT results by question difficulty showing Charts-of-Thought improvements across Easy, Moderate, and Hard questions for all three LLM models.}
    \label{fig:fig_4}
    \vspace{-10pt}
\end{figure}

Fig. \ref{fig:fig_4} shows LLM performance by question difficulty, as classified in the original VLAT paper. Questions are categorized as Easy, Moderate, or Hard based on their empirical difficulty for human participants.

\par All three models showed a pattern of decreasing performance as question difficulty increased, which aligns with human performance patterns. However, the performance gap between LLMs and humans widens with more difficult questions. Using the Charts-of-Thought approach, Claude-3.7-sonnet achieved remarkable results, correctly answering 94.1\% of Easy questions, 100\% of Moderate questions, and 88.2\% of Hard questions.

\par The standard prompting approach showed lower accuracy across all difficulty levels for all models. The most notable differences appeared in Hard questions, where Claude-3.7-sonnet's accuracy improved from 64.7\% with standard prompting to 88.2\% with Charts-of-Thought prompting. Similarly, GPT-4.5-preview's accuracy on Hard questions increased from 47.1\% to 70.6\%.

\subsection{Performance by Task Type}
\begin{figure}[!thb]
    \centering
    \includegraphics[width=0.45\textwidth, alt={Comparison}]{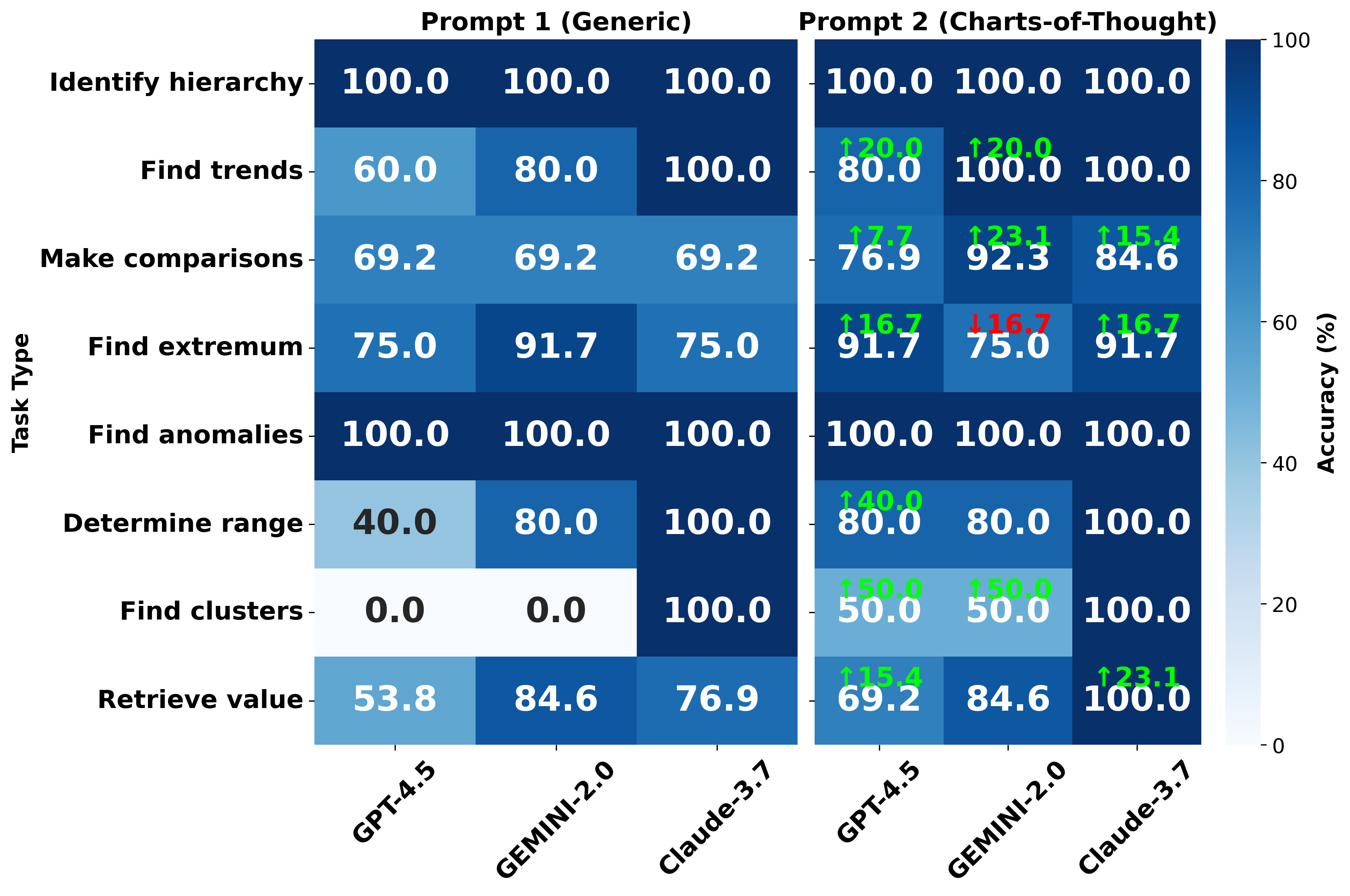}
    \caption{Modified VLAT results by task type comparing Generic and Charts-of-Thought prompting performance across eight analytical tasks.}
    \label{fig:fig_5}
\end{figure}

Fig. \ref{fig:fig_5} presents LLM performance broken down by analytical task type. The Modified VLAT includes eight task types: Retrieve Value, Find Extremum, Determine Range, Find Correlation/Trends, Make Comparisons, Find Anomalies, Find Clusters, and Identify Hierarchy.

\par With Generic prompting, all three models already achieve perfect accuracy (100\%) on the Identify Hierarchy and Find Anomalies tasks. In addition, Claude-3.7-sonnet also reaches 100\% accuracy on the Find Trends, Determine Range, and Find Clusters tasks. Charts-of-Thought prompting adds to these capabilities. Now Gemini-2.0-pro also reaches 100\% for Find Trends and Claude-3.7-sonnet also achieves 100\% for Retrieve Value (from 76.95\%). In fact, tasks involving precise value extraction and comparison
showed the most significant improvement with Charts-of-Thought prompting. This finding contrasts with previous research \cite{IN2}, which identified value retrieval as a particular weakness of LLMs in visualization tasks. 

For the other tasks we see general improvements for all models when the Chart-of-Thoughts prompting strategy is used (except for Find Extremum with Gemini-2.0-pro). The most dramatic improvement is for the Find Clusters task which was has a score of zero for GPT-4.5-preview and Gemini-2.0-pro with Generic prompting but 50\% with Charts-of-Thought prompting. This demonstrates that with structured prompting, modern LLMs can master many of the fundamental analytical tasks required for visualization interpretation.
%
%

%

\subsection{Performance by Visualization Type}
\begin{figure}[!thb]
    \centering
    \includegraphics[width=0.45\textwidth, alt={Comparison}]{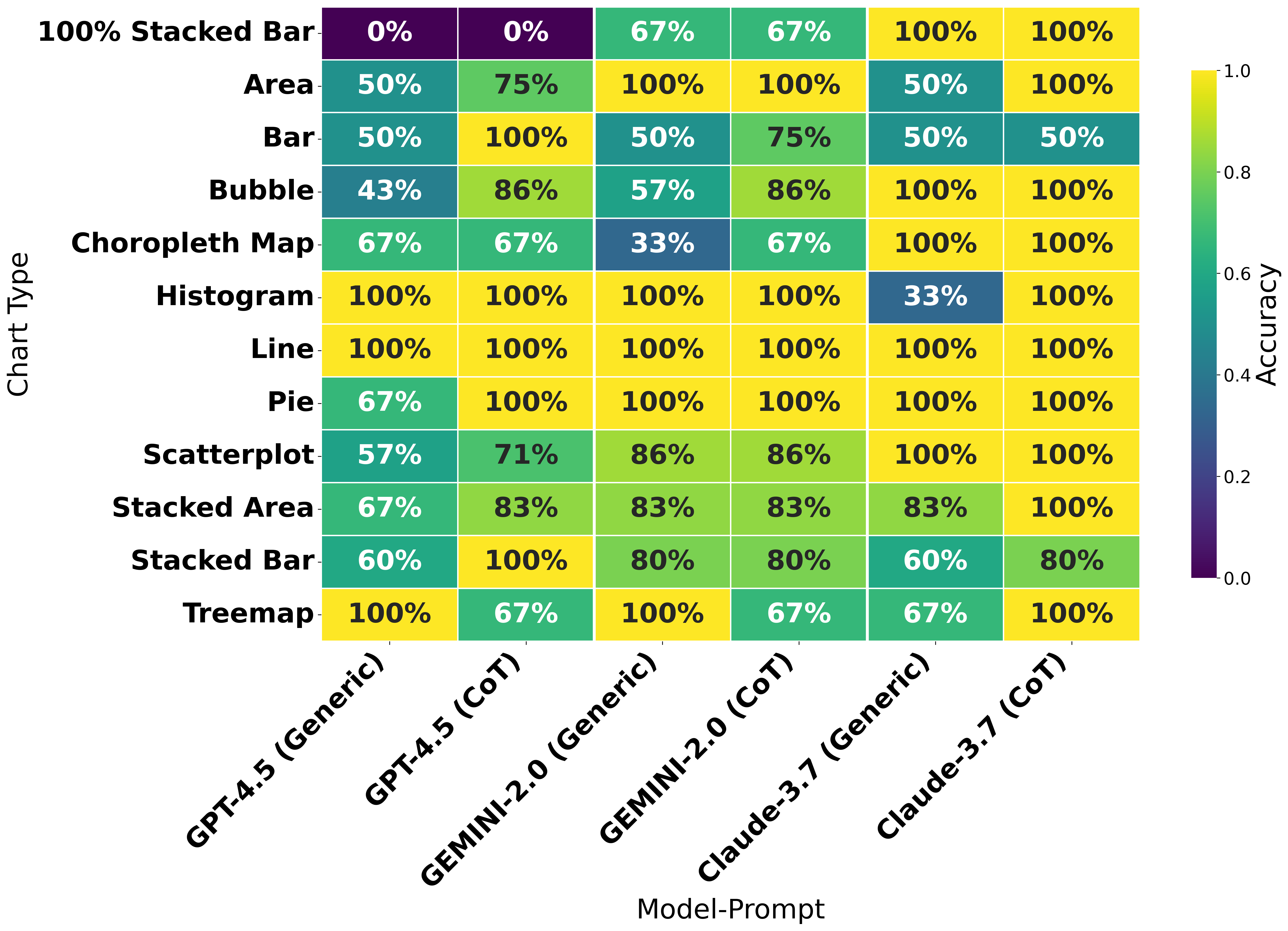}
    \caption{Modified VLAT results by chart type showing performance differences between prompting strategies across 12 visualization types.}
    \label{fig:fig_20}
\end{figure}

Fig. \ref{fig:fig_20} shows the LLM performance across the 12 visualization types included in the Modified VLAT. The results reveal interesting patterns in how different models handle the various visualization formats.

\par With Charts-of-Thought prompting, Claude-3.7-sonnet achieved 100\% accuracy on 10/12 visualization types: Stacked Bar Chart, Line Chart, Pie Chart, Histogram, Bubble Chart, Area Chart, Scatterplot, Stacked Bar chart, Choropleth Map, and Treemap. It performed less well on Stacked Bar Chart (slightly, 80\%) and Bar Chart (50\%).
\par GPT-4.5-preview showed substantial improvement on Bubble Chart questions, with accuracy increasing from 43\% with standard prompting to 86\% with Charts-of-Thought prompting. It also achieved perfect accuracy (100\%) on Line Chart, Pie Chart, Histogram, and Bar Chart-type questions (3/5 only with Charts-of-Thought prompting). Gemini-2.0-pro demonstrated its strongest performance on Area Chart, Line Chart, Pie Chart, and Histogram achieving 100\% accuracy on both with either prompting strategy. 
Yet, overall Gemini-2.0-pro benefited the least of the three models from Charts-of-Thought prompting and for the Treemap the prompting strategy had an adverse effect (also for GPT-4.5-preview). The most challenging visualization type for most models was the Bar Chart, with Claude-3.7-sonnet achieving only 50\% accuracy with both prompting strategies. This finding is surprising given the relative simplicity of bar charts and warrants further investigation.

\section{Experiments with the Original VLAT }
After testing our prompting strategies on modified VLAT visualizations, we evaluated the LLMs on the original VLAT charts to determine whether the improvements from Charts-of-Thought prompting would persist with standard benchmark materials. Based on our findings with the modified visualizations, we used only the Charts-of-Thought prompt (Prompt \#2) for all original VLAT experiments.

\subsection{Overall Performance}
Table \ref{tab:6} shows the performance of the three LLMs on the original VLAT questions compared to the human baseline from Lee et al. \cite{IN40} and GPT-4.o results reported by Bendeck and Stasko \cite{IN2}. We present both raw scores (out of 53 questions) and scores calculated using the VLAT scoring scheme.


%


\begin{table}[htbp]
\centering
\setlength{\tabcolsep}{4pt}
\caption{Results on original VLAT visualizations using CoT prompting}
\begin{tabular}{lcc>{\raggedleft\arraybackslash}p{1.5cm}}
\toprule
\textbf{Model} & \begin{tabular}[c]{@{}c@{}}\textbf{Mean Raw}\\\textbf{Score}\end{tabular} & \begin{tabular}[c]{@{}c@{}}\textbf{VLAT}\\\textbf{Score }\end{tabular} & \textbf{vs. Human} \\
\midrule
Human \cite{IN40} & 34.72 & 28.82 & baseline \\
GPT-4.o \cite{IN2} & 29.33 & 19.67 & \textcolor{red}{\textbf{-31.7\%}} \\
GPT-4.o (CoT) & 34.67 & 24.89 & \textcolor{red}{\textbf{-13.6\%}} \\
GPT-4.5 (CoT) & 42.00 & 34.33 & \textcolor{green!60!black}{\textbf{+19.1\%}} \\
GEMINI-2.0 (CoT) & 40.00 & 33.50 & \textcolor{green!60!black}{\textbf{+16.2\%}} \\
Claude-3.7 (CoT) & \textbf{51.00} & \textbf{50.17} & \textcolor{green!60!black}{\textbf{+74.1\%}} \\
\bottomrule
\end{tabular}

\label{tab:6}
\end{table}

\par In Table \ref{tab:6}, the models indicated with (CoT) use Charts-of-Thought prompting. We found that Claude-3.7-sonnet (CoT) achieved exceptional results on the original VLAT, correctly answering 51 out of 53 questions for a raw score of 96.2\%. Using the VLAT scoring scheme, Claude-3.7-sonnet (CoT) received a score of 50.17, far exceeding the human mean of 28.82 reported in the original study.
\par Both GPT-4.5-preview (CoT) and Gemini-2.0-pro (CoT) also performed well above the human baseline. GPT-4.5-preview (CoT) achieved a VLAT score of 34.33, while Gemini-2.0-pro (CoT) scored 33.50, both significantly higher than the human mean of 28.82 and the GPT-4.o score of 19.67 reported by Bendeck and Stasko \cite{IN2}. 

In addition, we also ran Charts-of-Thoughts prompting on the original GPT-4.o model used by Bendeck and Stasko \cite{IN2}. We found that while this GPT-4.o (CoT) variant improved in accuracy, indicating that the structured prompting brought benefits ($\sim$25\%), it still could not beat the human baseline. This indicates the impressive performance gains in visual literacy of the more recent LLMs overall, which we further substantially enhanced with structured prompting as demonstrated in the previous sections.

\subsection{Performance by Question Difficulty}
Table \ref{tab:7} breaks down LLM performance by question difficulty, as classified in the original VLAT paper.


%

\definecolor{lightgreen}{RGB}{204,255,204}
\definecolor{mediumgreen}{RGB}{102,204,102}
\definecolor{darkgreen}{RGB}{0,153,0}

\begin{table}[htbp]
\centering
\setlength{\tabcolsep}{4pt}
\caption{Original VLAT Results by Question Difficulty}
\setlength{\tabcolsep}{5pt}
\begin{tabular}{|l|>{\columncolor{white}}c|>{\columncolor{white}}c|>{\columncolor{white}}c|>{\columncolor{white}}c|}
\hline
\rowcolor{gray!20} Difficulty & GPT-4.o \cite{IN2} & \begin{tabular}[c]{@{}c@{}}GPT-\\4.5\\preview\end{tabular} & \begin{tabular}[c]{@{}c@{}}GEMINI-\\2.0\\pro\end{tabular} & \begin{tabular}[c]{@{}c@{}}Claude-\\3.7\\sonnet\end{tabular} \\
\hline
Easy & \cellcolor{lightgreen}70.6\% & \cellcolor{mediumgreen}94.1\% & \cellcolor{lightgreen}82.4\% & \cellcolor{darkgreen}100.0\% \\
 & (12/17) & (16/17) & (14/17) & (17/17) \\
\hline
Moderate & \cellcolor{gray!30}52.6\% & \cellcolor{lightgreen}63.2\% & \cellcolor{lightgreen}63.2\% & \cellcolor{darkgreen}100.0\% \\
 & (10/19) & (12/19) & (12/19) & (19/19) \\
\hline
Hard & \cellcolor{gray!15}35.3\% & \cellcolor{lightgreen}58.8\% & \cellcolor{gray!30}52.9\% & \cellcolor{mediumgreen}88.3\% \\
 & (6/17) & (10/17) & (9/17) & (15/17) \\
\hline
\end{tabular}

\label{tab:7}
\end{table}

\par Claude-3.7-sonnet with Charts-of-Thought prompting demonstrated superior performance across all difficulty levels, answering 100\% of Easy and Moderate questions correctly. Even for Hard questions, Claude-3.7-sonnet achieved 88.3\% accuracy, more than double the 35.3\% reported for GPT-4.o by Bendeck and Stasko \cite{IN2}.
\par GPT-4.5-preview with Charts-of-Thought prompting showed substantial improvement over GPT-4.o, particularly on Easy questions (94.1\% vs. 70.6\%) and Hard questions (58.8\% vs. 35.3\%). Gemini-2.0-pro with Charts-of-Thought prompting performed at a similar level to GPT-4.5-preview, with slightly lower accuracy on Easy questions but comparable results on Moderate and Hard questions.
\subsection{Performance by Task Type}
\begin{figure}[!thb]
    \centering
    \includegraphics[width=0.45\textwidth, alt={Comparison}]{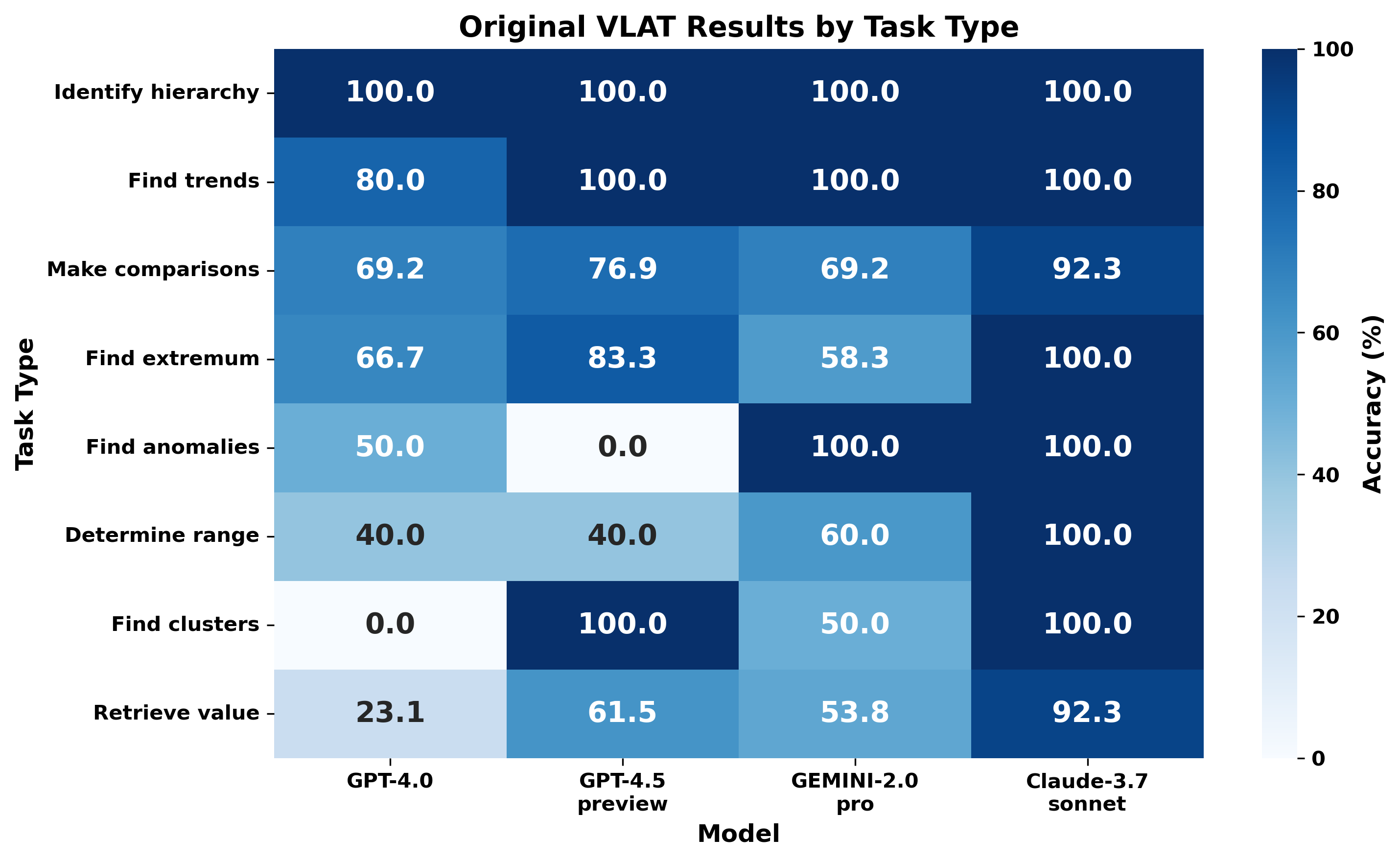}
    \caption{Original VLAT results by task type comparing Generic and Charts-of-Thought prompting performance across eight analytical tasks.}
    \label{fig:fig_11}
    \vspace{-10pt}
\end{figure}

Fig. \ref{fig:fig_11} presents LLM performance broken down by analytical task type for the original VLAT. All variants except the baseline GPT-4.o use Charts-of-Thought prompting. 


%

%
\par The most notable improvement was in the Retrieve Value task, where Claude-3.7-sonnet achieved 92.3\% accuracy compared to just 23.1\% for GPT-4.o. This contradicts previous findings that suggested value retrieval was a particular weakness of LLMs in visualization tasks.
\par Claude-3.7-sonnet reached 100\% accuracy on five task types: Identify Hierarchy, Find Trends, Find Extremum, Find Anomalies, Determine Range, and Find Clusters. GPT-4.5-preview excelled at Find Trends (100\%) and Find Clusters (100\%), while Gemini-2.0-pro performed best on Find Trends (100\%) and Find Anomalies (100\%).
\par The most challenging task across all models was Determine Range, with GPT-4.5-preview achieving only 40.0\% accuracy. However, Claude-3.7-sonnet still managed perfect performance on this task.
\subsection{Performance by Chart Type}
\begin{figure}[!thb]
    \centering
    \includegraphics[width=0.45\textwidth, alt={Comparison}]{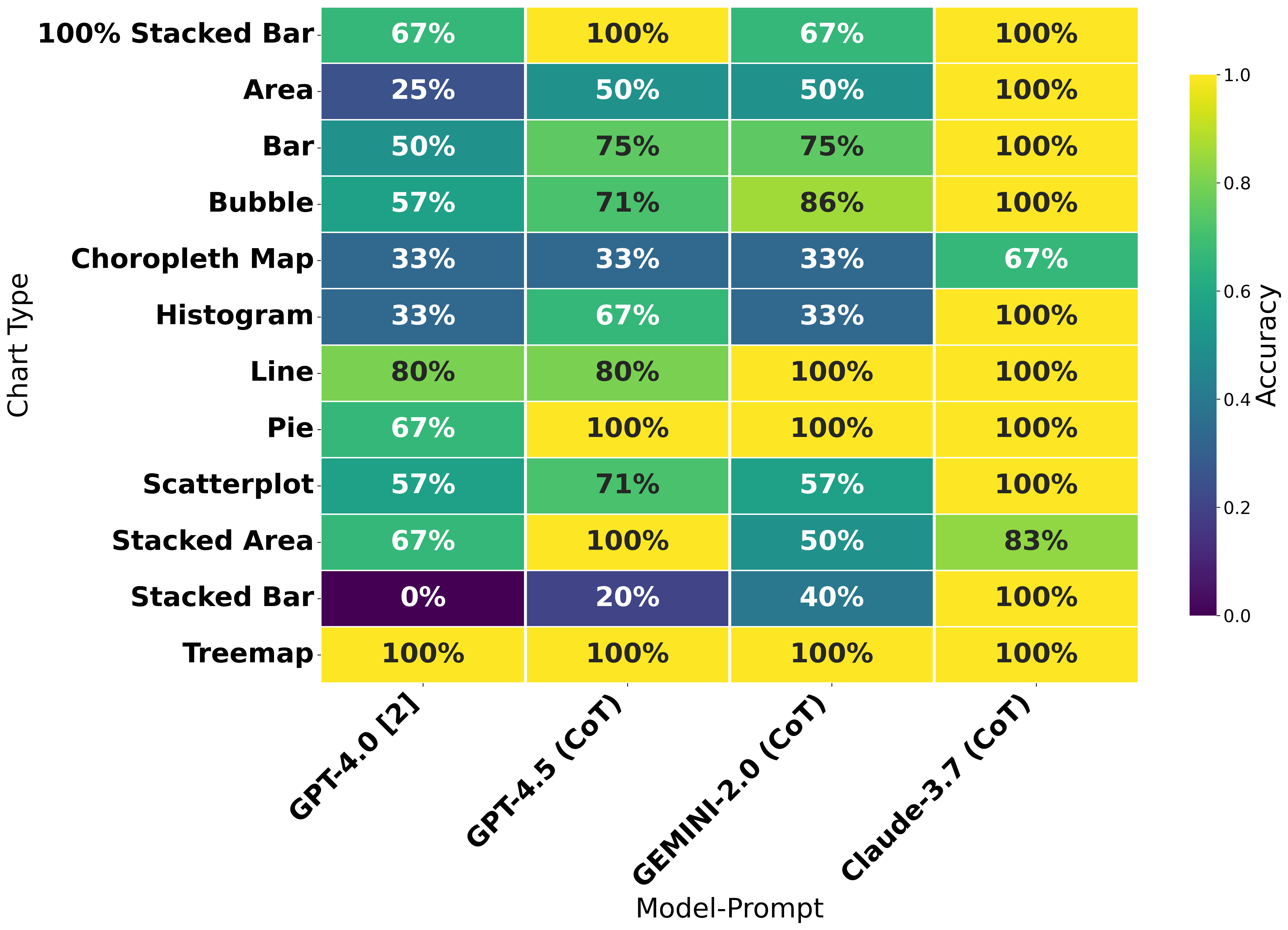}
    \caption{Original VLAT results by chart type showing performance differences between prompting strategies across 12 visualization types.}
    \label{fig:fig_30}
        \vspace{-15pt}
\end{figure}

Fig. \ref{fig:fig_30} shows LLM performance across the 12 visualization types included in the original VLAT. All variants except the baseline GPT-4.o use Charts-of-Thought prompting.

\par Claude-3.7-sonnet achieved 100\% accuracy on 10 of the 12 visualization types, with slightly lower performance on Choropleth Maps (67\%) and Stacked Area Charts (83\%). This represents a dramatic improvement over GPT-4.o, which struggled with several chart types, most notably Stacked Bar Charts (0.0\%) and Area Charts (25.0\%).
\par GPT-4.5-preview performed well on Treemap (100\%), Pie Chart (100\%), Stacked Area Chart (100\%), and 100\% Stacked Bar Chart (100\%), but struggled with Choropleth Maps (33\%) and Stacked Bar Charts (20.0\%). Gemini-2.0-pro showed its strongest performance on Treemap (100\%), Line Chart (100\%), and Pie Chart (100\%).

\par The most challenging visualization type for all models was the Choropleth Map, with even Claude-3.7-sonnet achieving only 67\% accuracy. This suggests that geographical visualizations remain difficult for LLMs to interpret correctly.

\subsection{Comparison with Modified VLAT Results}
The performance patterns on the original VLAT largely mirrored those observed on our modified VLAT presented earlier. Claude-3.7-sonnet consistently outperformed the other models on both sets, and the Charts-of-Thought prompting strategy proved effective for all LLMs tested.
\par One key observation is that Claude-3.7-sonnet performed only slightly better on the original VLAT (50.17) than on our modified version (49.44). This suggests that the model's strong performance is not due to prior exposure to the original VLAT materials. 
%
\par Conversely, both GPT-4.5-preview and Gemini-2.0-pro showed slightly lower scores on the original VLAT compared to the modified version, but their performance remained well above the human baseline and previous GPT-4.o results.

\section{Chart Question Answering}
To evaluate our Charts-of-Thought prompting strategy beyond the VLAT, we tested its effectiveness on chart question answering (CQA) tasks. Based on our findings from the VLAT experiments, we focused on testing only Claude-3.7-sonnet, which consistently demonstrated superior visualization literacy capabilities across both modified and original VLAT tests. Claude-3.7-sonnet not only achieved the highest overall scores (VLAT score of 49.44 on modified and 50.17 on original tests), but also showed exceptional performance across various visualization types and analytical tasks. Given these results, we compared Claude-3.7-sonnet's performance using our charts-of-thought prompting technique against the chart question answering system developed by Kim et al. \cite{IN32}.

\subsection{Experimental Setup}
We used the CQA dataset from Kim et al. \cite{IN32}, which consists of 629 questions across 47 bar charts (32 simple, 8 grouped, 7 stacked) and 5 line charts. Based on our findings from the VLAT experiments, we focused on testing only Claude-3.7-sonnet, which showed the strongest visualization literacy capabilities. Following Bendeck and Stasko \cite{IN2}, who found that LLMs perform well when provided with the underlying data but struggle without it, we chose to test only the more challenging condition: answering questions without providing the underlying data. This approach better measures the model's visualization literacy rather than its question-answering capabilities with structured data.

\par We applied our Charts-of-Thought prompt to all questions in the dataset, running the experiments three times to ensure reliability. For numerical answers, we considered responses correct if they were within 5\% of the actual value, consistent with prior work \cite{IN52}.

\subsection{Results}
\begin{figure}[!thb]
    \centering
    \includegraphics[width=0.45\textwidth, alt={Comparison}]{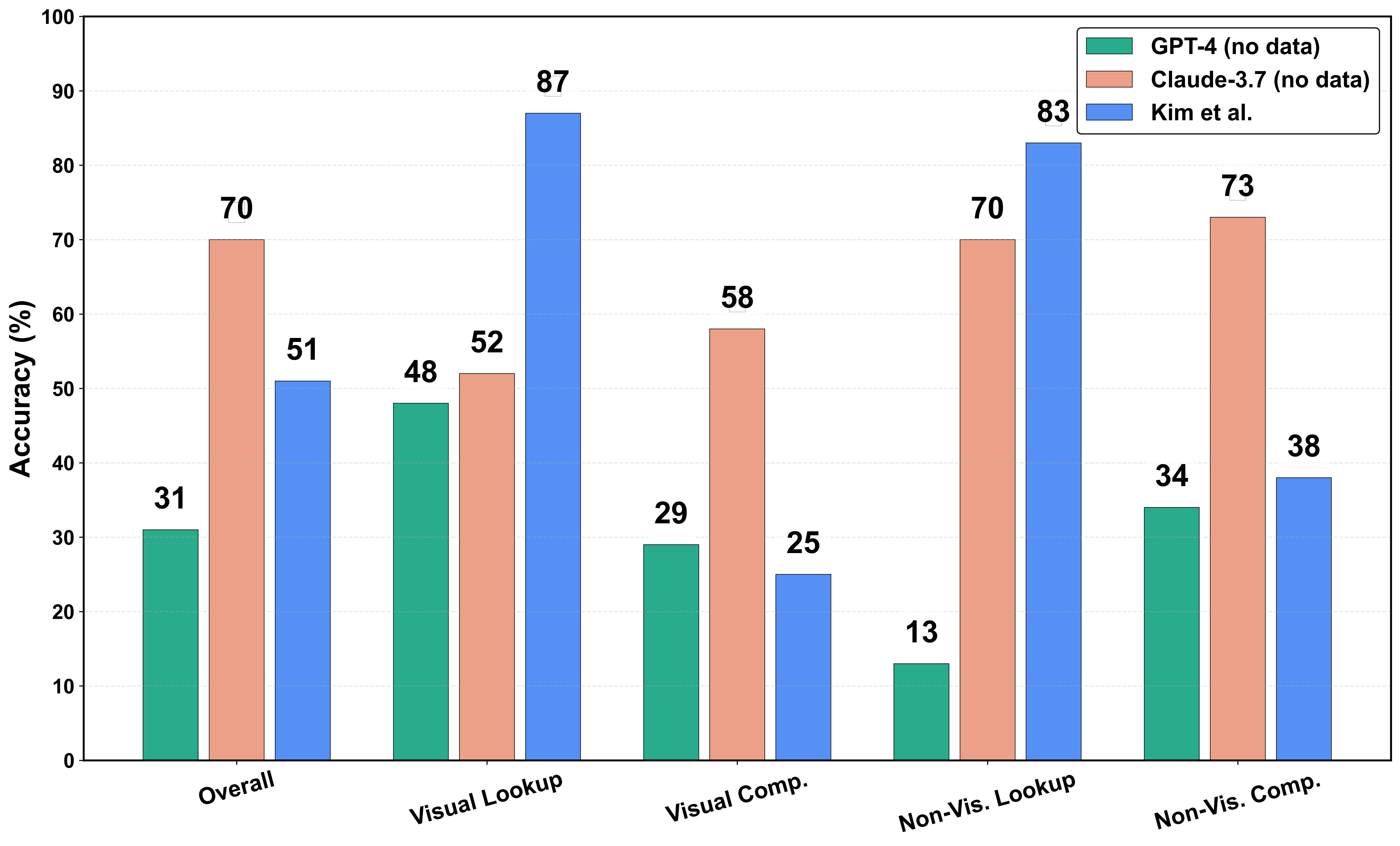}
    \caption{CQA accuracy of Claude-3.7-sonnet with Charts-of-Thought prompting compared to GPT-4 without data and Kim et al.'s system}
    \label{fig:fig_3}
\end{figure}

Fig. \ref{fig:fig_3} shows Claude-3.7-sonnet's performance compared to the baseline system from Kim et al. \cite{IN32} and the results reported for GPT-4 without data in Bendeck and Stasko \cite{IN2}.

\subsection{Overall Performance}
Claude-3.7-sonnet with Charts-of-Thought prompting achieved an overall accuracy of 70\% on the CQA dataset, substantially outperforming both GPT-4 without data (31\%) and Kim et al.'s specialized CQA system (51\%). This result demonstrates the effectiveness of our structured prompting approach in enhancing LLMs' ability to extract and reason about data from visualizations.

\subsection{Performance by Question Type}
We analyzed performance across four question categories: visual lookup, non-visual lookup, visual compositional, and non-visual compositional questions. Claude-3.7-sonnet performed best on non-visual compositional questions (73\%), which require multiple operations on data values without referencing specific visual features. It surpassed Kim et al.'s system (38\%) by a wide margin and doubled GPT-4's performance (34\%).

\par For visual compositional questions, which reference visual elements while requiring multiple operations, Claude-3.7-sonnet achieved 58\% accuracy, compared to 25\% for Kim et al.'s system and 29\% for GPT-4. On non-visual lookup questions, which involve simple value retrieval without visual references, Claude-3.7-sonnet reached 70\% accuracy, compared to 13\% for GPT-4 and 83\% for Kim et al.'s system. This represents the second category where our approach did not exceed the baseline system. Visual lookup questions, which reference specific visual features for value retrieval, proved most challenging for all systems. Claude-3.7-sonnet achieved 52\% accuracy, outperforming GPT-4 (48\%) but Kim et al.'s system scored (87\%).

\subsection{Performance by Task Type}
Table \ref{tab:10} shows Claude-3.7-sonnet and GPT-4's performance on various analytical tasks within the CQA dataset. Claude-3.7-sonnet performed best on tasks requiring computation of derived values (72\%) and make comparisons (72\%), demonstrating its ability to perform complex analytical tasks on visualization data. These results align with our VLAT findings, where structured prompting improved performance on tasks requiring multi-step reasoning. Multiple tasks showed 69\% accuracy, while Lookup tasks reached 65\%. Find extrema tasks, though improved from previous LLM results, remained relatively challenging at 62\% accuracy.

\definecolor{lowperf}{RGB}{255,200,200}
\definecolor{midperf}{RGB}{255,235,173}
\definecolor{highperf}{RGB}{198,239,206}
\newcommand{\perf}[1]{%
\ifnum#1<25\cellcolor{lowperf}#1\%%
\else\ifnum#1<50\cellcolor{midperf}#1\%%
\else\cellcolor{highperf}#1\%%
\fi\fi
}
\begin{table}[htbp]
\centering
\setlength{\tabcolsep}{4pt}
\caption{Claude-3.7-sonnet and GPT-4's performance on various analytical tasks within the CQA dataset}
\setlength{\tabcolsep}{4pt}
\begin{tabular}{lccc}
\toprule
\rowcolor{gray!10}
\textbf{Task} & \textbf{\#Questions} & \multicolumn{2}{c}{\textbf{Accuracy w/ data}} \\
\rowcolor{gray!10}
& & \textbf{GPT-4.o} & \textbf{Claude-3.7} \\
\midrule
Compute derived value & 125 & \perf{7} & \perf{72} \\
Lookup & 193 & \perf{23} & \perf{65} \\
Find extrema & 267 & \perf{52} & \perf{62} \\
Make comparisons & 25 & \perf{44} & \perf{72} \\
Multiple & 70 & \perf{37} & \perf{69} \\
\bottomrule
\end{tabular}

\label{tab:10}
\end{table}

\subsection{Analysis of Error Cases}
We examined cases where Claude-3.7-sonnet failed to provide correct answers. 
Three main types of error patterns emerged:

\begin{itemize} 
    \item \textbf{\textit{Color interpretation errors:}} 
    Similar to findings in the VLAT experiments, Claude-3.7-sonnet sometimes misinterpreted color encodings in charts with multiple colors, particularly in stacked and grouped bar charts. For example, when asked about the proportion of a specific colored segment in a stacked bar, the model sometimes confused the color references.
    \item \textbf{\textit{Axis misinterpretation:}} 
    In some cases, Claude-3.7-sonnet struggled with unusual axis scales or non-zero baselines, leading to incorrect readings. The Charts-of-Thought prompt reduced these errors compared to generic prompting but did not eliminate them.
    \item \textbf{\textit{Complex calculation errors:}} 
    For questions requiring multiple calculation steps, occasional arithmetic errors appeared in the model's reasoning chain, despite the structured approach.  
\end{itemize}

\section{Discussion}
Our study reveals that modern Large Language Models (LLMs) can achieve remarkable visualization literacy when guided through structured analytical processes. Through our Charts-of-Thought prompting strategy, we have demonstrated that these models can not only match but substantially exceed human performance on standardized visualization literacy assessments. This section discusses our findings in relation to our research questions and examines the implications for the visualization community.

\subsection{Results Summary}
\begin{itemize}

    \item \textbf{RQ1: LLM Visualization Literacy With Structured Guidance}
Our experiments clearly show that modern multimodal LLMs can achieve and surpass human-level visualization literacy when guided through a structured analytical process. Claude-3.7-sonnet achieved a VLAT score of 50.17, far exceeding the human baseline of 28.82. This finding challenges the prevailing view from previous research that LLMs lack sufficient visualization literacy for evaluation tasks. The performance of Claude-3.7-sonnet is particularly noteworthy, as it correctly answered 51 out of 53 questions on the original VLAT using our Charts-of-Thought approach. This level of accuracy suggests that modern LLMs possess strong capabilities for interpreting visual data when properly guided. Both GPT-4.5-preview and Gemini-2.0-pro also performed well above human baselines, demonstrating that this capability extends across multiple state-of-the-art models.

    \item \textbf{RQ2: Impact of Charts-of-Thought Prompting}
Our Charts-of-Thought prompting strategy consistently improved performance across all tested models. GPT-4.5-preview showed the most dramatic improvement with a 21.8\% increase in raw score when using structured prompting compared to standard instructions. Gemini-2.0-pro and Claude-3.7-sonnet also demonstrated significant improvements of 9.4\% and 13.5\% respectively. These improvements were consistent across both our modified VLAT materials and the original benchmark visualizations, indicating that the structured approach helps LLMs extract and analyze data more effectively regardless of their prior exposure to the specific charts. The Charts-of-Thought approach appears to align with how humans naturally process visualizations—first extracting data points, then organizing this information, and finally performing analysis to answer questions.

    \item \textbf{RQ3: LLM Performance Across Visualization Types and Tasks}
Our analysis revealed varying performance patterns across visualization types and analytical tasks. Claude-3.7-sonnet demonstrated exceptional versatility, achieving 100\% accuracy on seven visualization types with the Charts-of-Thought approach. More complex visualizations like choropleth maps proved challenging for all models, with even Claude-3.7-sonnet achieving only 66.7\% accuracy on these charts. For analytical tasks, all three models achieved perfect accuracy on the Identify Hierarchy task with Charts-of-Thought prompting. Claude-3.7-sonnet also reached 100\% accuracy on Find Trends, Find Anomalies, Determine Range, and Find Clusters tasks. Value retrieval, which prior research identified as a particular weakness of LLMs, saw dramatic improvement with our approach, with Claude-3.7-sonnet achieving 100\% accuracy on the Retrieve Value task on the modified VLAT and 92.3\% on the original VLAT. The Chart Question Answering experiments further demonstrated the effectiveness of our approach across different analytical tasks. Claude-3.7-sonnet performed best on tasks requiring computation of derived values (72\%) and making comparisons (72\%), showing that structured prompting enhances LLMs' ability to perform complex analytical operations on visualization data.
 \item \textbf{RQ4: Prior Knowledge vs. Visualization Information}
Our experiments with modified VLAT materials helped assess whether LLMs rely on prior knowledge or information presented in the visualizations. The fact that models performed well on our modified charts with randomized data suggests they are not simply recalling answers from their training data but are actually interpreting the visual information. However, we did observe that LLMs performed slightly better on original VLAT materials than on our modified versions, suggesting some benefit from potential prior exposure. This effect was most pronounced for Claude-3.7-sonnet, which scored 50.17 on the original VLAT compared to 49.44 on our modified version. This difference is relatively small and does not detract from the overall finding that LLMs can effectively interpret novel visualizations.

\end{itemize}


%
%
\subsection{Limitations and Future Work}
Despite the impressive performance of LLMs on visualization literacy tasks, several limitations remain. Color interpretation remains a significant challenge for LLMs, even with our Charts-of-Thought approach. This limitation was most apparent in visualizations where color encoding carries critical information, such as choropleth maps (66.7\% accuracy), stacked bar charts (80\% accuracy), and grouped bar charts. We observed that models struggled with three specific color-related tasks: distinguishing between similar hues, interpreting color gradients in sequential color scales, and maintaining consistency when referencing the same color across multiple questions. For example, Claude-3.7-sonnet sometimes misattributed data values when asked about specific colored segments in stacked visualizations. This challenge becomes more pronounced as visualization complexity increases, suggesting that future prompting strategies need to include more explicit color verification steps. The consistent underperformance on color-dependent visualizations across all tested models indicates this may be a fundamental limitation in current multimodal LLMs rather than a model-specific issue. Future work should explore specialized prompting strategies to address this persistent challenge.



Building on our results, we identify four promising research directions. A first such direction is to develop specialized Charts-of-Thought variants for complex visualization types not covered in VLAT, such as network graphs, parallel coordinates, and Sankey diagrams. These visualizations encode relationships rather than values, requiring different extraction and analysis steps. 

Second, future work should explore interactive Charts-of-Thought approaches. Instead of using fixed prompts, systems could adapt the prompting sequence based on visualization complexity and initial model responses. This would create more robust interpretation capabilities for unusual chart types or data distributions. 

Third, another impactful research area would be to investigate domain-specific Charts-of-Thought prompts for specialized visualizations in fields like genomics, astronomy, and finance. These fields use visualization conventions that require background knowledge to interpret properly, presenting unique challenges for LLMs. 


Finally, although our results show that LLMs can interpret static visualizations with high accuracy, their ability to evaluate visual design choices or provide design recommendations remains largely unexplored. Future work should assess whether LLMs can not only interpret but also critique visualizations based on established design principles.

\subsection{Implications for Visualization Research and Practice}

Our findings have implications for the visualization community. First, they show that multimodal LLMs, when properly prompted, can serve as effective tools for interpreting and evaluating visualizations. This opens the door to automating assessments of visualization designs, reducing the time and cost of human evaluation. Second, the Charts-of-Thought approach offers a framework for improving LLM performance on visualization tasks, with applications across multiple domains. For example, data dashboards could integrate LLMs to provide interpretation help, making charts more accessible to non-specialists. Education platforms could offer feedback to students learning visualization design. Accessibility tools could generate alt-text descriptions for screen readers, aligning with the insights that sighted users gain. Third, the performance of LLMs on tasks such as value retrieval and trend identification suggests they could power natural language interfaces for data exploration. With verification mechanisms, this could democratize access to visualization tools for users without specialized training. Finally, our results establish a benchmark for LLM visualization literacy and highlight the value of structured prompting strategies for visual interpretation tasks. As LLMs continue to advance, we anticipate a growing role for these models in how we create, evaluate, and interact with data visualizations.

\subsection{Ethical Considerations}
Despite these benefits, several concerns remain. In high-stakes domains like healthcare, LLMs should supplement, not replace, human analysis. Hallucination risk persists—models may generate plausible but incorrect interpretations, especially with complex visualizations. Organizations should implement verification steps when using these systems. Over-reliance on LLMs could lead to skill loss among professionals if they routinely defer to automation. Finally, visualization designers might begin to optimize for AI readability over human comprehension, which would undermine the purpose of data visualization.

\section{Conclusion}
Our study shows that modern multimodal LLMs can exceed human performance on visualization literacy tasks when guided through the Charts-of-Thought approach. Claude-3.7-sonnet achieved a VLAT score of 50.17, far above the human baseline of 28.82, with similar improvements for all tested models. This structured prompting technique consistently improved performance across visualization types and analytical tasks, particularly for value retrieval and comparison operations that previously challenged LLMs. These findings establish new benchmarks for machine interpretation of visualizations and suggest practical applications in automated evaluation, accessibility, and data analysis. Future work should address remaining challenges with color interpretation, geographical visualizations, and developing automated prompting strategies for more complex visualization formats.

\section*{\textbf{\textsf{\textsc{Supplemental Materials}}}}

To support reproducibility and future research, we provide comprehensive supplementary materials, including all codes, stimuli, and results evaluated in this study. These materials are available as a .zip file through the PCS Submission System and are also publicly accessible at \textcolor{blue}{https://github.com/vhcailab/Charts-of-Thought}. The description and location of all supplemental materials are provided as a separate document named "Supplemental Materials Details.pdf" inside the zipped folder.

 \acknowledgments{
This material is based upon work supported by the National Science Foundation under Grant No. NRT-HDR 2125295. Any opinions, findings, and conclusions or recommendations expressed in this material are those of the author(s) and do not necessarily reflect the views of the National Science Foundation.
 }

\bibliographystyle{abbrv-doi-hyperref}

\bibliography{template}

\appendix 

\end{document}